\documentclass[12pt,a4paper,twoside]{article}
\usepackage[utf8]{inputenc}
\usepackage[english]{babel}
\usepackage[margin=3cm]{geometry}

\usepackage{hyperref}
\usepackage{graphicx}
\usepackage{times}
\usepackage{fancyhdr}
\usepackage{epsfig}
\usepackage{multirow}
\usepackage{amssymb,amsfonts,amsmath}
\usepackage{nameref}
\usepackage{csquotes}
\usepackage{xcolor}
\usepackage{pdfpages}

\usepackage{graphicx}

\newcommand{\op}[1]{\mathbf{#1}}
\usepackage{color}

\let\Re\relax

\DeclareMathOperator{\Re}{\mathrm{Re}}

\usepackage[normalem]{ulem}
\usepackage{soul}

\usepackage[footnotesize,bf]{caption}

\DeclareCaptionLabelFormat{naturelabel}{\textbf{Fig.#2$\mid$}}
\usepackage{fancyhdr}

\begin{document}

\pagestyle{fancy}
\setlength{\headheight}{16pt}
\fancyhead{}
	

\renewcommand{\multirowsetup}{\centering}

\thispagestyle{empty} 

\begin{center}
	{\Large\bfseries\sffamily
	Estimates of the reproduction ratio from epidemic surveillance may be biased\\in spatially structured populations
\\
	}
	\vspace{1cm}\large
	{Piero Birello\textsuperscript{1},
        Michele Re Fiorentin\textsuperscript{2},
	 Boxuan Wang\textsuperscript{1},\\
	 Vittoria Colizza\textsuperscript{1}, and
	Eugenio Valdano\textsuperscript{1,*}.
 	}\\
	\vspace{1.1cm}
	{\footnotesize \textit{
		\textsuperscript{1}Sorbonne Universit\'{e}, INSERM, Institut Pierre Louis d'Epid\'{e}miologie et de Sant\'{e} Publique,\\ F75012, Paris, France.\\
		\textsuperscript{2}Department of Applied Science And Technology (DISAT), Politecnico di Torino, 10129, Torino, Italy.
\\
		\textsuperscript{*} Corresponding author} \texttt{eugenio.valdano@inserm.fr}
	\\
	}
\end{center}

\vspace{1.1cm}



\begin{center}\sffamily\large\bfseries
    Abstract
\end{center}
Accurate estimates of the reproduction ratio are crucial to project infectious disease epidemic evolution and guide public health response. Here, we prove that estimates of the reproduction ratio based on inference from surveillance data can be inaccurate if the population comprises spatially distinct communities, as the space-mobility interplay may hide the true epidemic evolution from surveillance data. Consequently, surveillance may underestimate the reproduction ratio over long periods, even mistaking growing epidemics as subsiding. To address this, we use the spectral properties of the matrix describing the spatial epidemic spread to reweigh surveillance data. We propose a correction that removes the bias across all epidemic phases. We validate this correction against simulated epidemics and use COVID-19 as a case study. However, our results apply to any epidemic where mobility is a driver of circulation. Our findings will help improve epidemic monitoring and surveillance and inform strategies for public health response.

\newpage

\begin{center}\sffamily\large\bfseries
    Main text
\end{center}



\noindent The reproduction ratio $R$ is arguably the most used indicator to monitor the trend in the evolution of an infectious disease epidemic.
$R$ is the average number of secondary infections that each case generates: When it is larger than one, the epidemic wave is growing; when instead it is lower than one, it is subsiding~\cite{keeling_modeling_2007,nishiura_effective_2009}.
The reproduction ratio also measures the effectiveness of public health interventions, whose overarching goal is to bring an unconstrained epidemic ($R>1$) below the epidemic threshold of $R=1$: Accurately estimating the reproduction ratio is thus necessary to ascertain the current epidemic evolution, predict short-term trends, perform scenario analysis and plan public health action~\cite{wallinga_optimizing_2010,ridenhour_unraveling_2018,thompson_control_2018,dhillon_getting_2020,pan_association_2020}.
The standard way to measure $R$ is to infer it from data coming from epidemiological surveillance~\cite{wallinga_how_2007,davoudi_early_2012,obadia_r0_2012,cori_new_2013,thompson_improved_2019}.
These data may be timelines of detected cases or their proxies, like hospitalizations or deaths, and this approach applies to diseases spanning radically different epidemiology, transmission routes and burden, like influenza~\cite{biggerstaff_estimates_2014,thompson_global_2022}, measles~\cite{guerra_basic_2017}, COVID-19~\cite{li_temporal_2021}, Ebola~\cite{maganga_ebola_2014}, cholera~\cite{mukandavire_estimating_2011}, dengue~\cite{codeco_estimating_2018}, malaria~\cite{routledge_estimating_2018}.
The resulting surveillance-based estimates of $R$ are routinely used to design interventions~\cite{noauthor_introducing_2021}:
Notwithstanding, we argue in this study that surveillance data may lead to biased estimates of the reproduction ratio in spatially structured populations, where geographically distinct communities (e.g., cities) are connected though human mobility.
Heterogeneities in the contact network shape the way epidemics spread~\cite{pastor-satorras_epidemic_2015}.
Here, we will show that those arising from the complex interplay between spatial transmissibility patterns and the mixing network driven by human mobility hide the true dynamic structure of the epidemic process from population-level surveillance data.
This mirrors the nature of most mathematical models of epidemic spread: they integrate space and spatial data at high resolution~\cite{hufnagel_forecast_2004,balcan_phase_2011,pastor-satorras_epidemic_2015,soriano-panos_spreading_2018,gomez-gardenes_critical_2018,chang_mobility_2020}, but they find it harder to do the reverse, which is extracting high-resolution information from limited and coarse-grained surveillance data in the absence of knowledge of the underlying spatial dynamics~\cite{coletti_shifting_2018,scarpino_predictability_2019,castro_turning_2020}.
Crucially, this means that inference on surveillance data may either overestimate or underestimate it over long periods.
This is of great public health relevance: measuring for instance a reproduction ratio below one when the true value is above would falsely signal that the epidemic is under control.
Here, we study this bias, identify its origin and compute its magnitude.
Then, we propose a correction to case incidence data that removes this bias and ensures that surveillance-based estimates of the reproduction ratio consistently give the true reproduction ratio of the epidemic. 
Our theoretical findings apply to any epidemic featuring relatively short generation time and for which mobility is a contributing factor in shaping its circulation within and across communities.
This covers some of the global health threats that are being worst affected by climate change and demographic trends:
viruses responsible for respiratory infections --~including SARS-CoV-2 and influenza~--~\cite{li_trends_2022}, vector-borne pathogens --~including the arboviruses dengue, chikungunya, Zika~\cite{messina_current_2019,romanello_2022_2022}, and emergence events of new viruses or new viral strains~\cite{carlson_climate_2022}.
To test and illustrate our findings, we use the French COVID-19 epidemic (see Fig.~\ref{fig:fig1}) before the advent of vaccination as a case study. 



\section*{Theoretical formalism}

The Galton--Watson branching process is a customary framework to model epidemic spread~\cite{Watson,lloyd-smith_superspreading_2005,hellewell_feasibility_2020}. Let $I(0)$ be the initial number of infections, $I(1)$ the expected number of infections that the initial cases generate, and, generally, let $I(t)$ be the expected number of infections in the $t$-th generation. By definition of the reproduction ratio, we have that $I(t)=R I(t-1)$, which implies that $I(t)=R^{t}I(0)$.
This equation means that the number of infections grows exponentially if $R>1$.
In any real outbreak other factors, like acquired immunity, seasonal effects or public-health interventions, will at some point curb this exponential growth by changing the value of $R$.
Notwithstanding, we may assume $R$ to be fairly constant either in the early phase of an outbreak, when those effects have not yet kicked in, or when the timescale at which immunity and mixing change is much longer than epidemic evolution~\cite{kucharski2016effectiveness,de_meijere_attitudes_2023}.

In the case of a population composed of $N$ spatial communities, we may define the vector $\op{I}(t)\in\mathbb{R}^N$, whose component $I(t)_{i}$ is the number of infections in generation $t$ and community $i$. Likewise, the {\itshape reproduction operator} $\op{R}\in\mathbb{R}^{N,N}$ encodes, in its component $R_{ij}$, the average number of infections generated among the residents of community $i$, by a case belonging to community $j$ \cite{Susswein}. Its definition thus matches that of a spatially-structured {\itshape next-generation matrix}~\cite{diekmann_definition_1990,nishiura_pros_2010}. However, unlike the standard next-generation matrix, it can be built at any phase of the epidemic and not only close to the disease-free state, and with no need of assuming a specific transmission model.
The definition of $\op{R}$, and the results that we are going to derive from it, thus applies to any epidemic and disease.
The specific parametrization of $\op{R}$ will instead depend on the specific transmission dynamics and natural history of the disease: for directly-transmitted diseases $\op{R}$ typically depends on mixing patterns among communities~\cite{mazzoli_projecting_2021}; for vector-borne diseases the local abundance of the host vectors, modulating the effective transmissibility, needs to be factored in, too~\cite{messina_current_2019,jourdain_importation_2020}.
The expected epidemic evolution then follows the equation
\begin{equation}
    \op{I}(t) = \op{R}^t\op{I}(0).
    \label{eq:Ievol}
\end{equation}
$\op{I}(t)$ encodes both the total number of infections in the population in generation $t$ and its spatial distribution. We define the former as the number $I_{tot}(t)=\sum_i I(t)_{i}$ and the latter as the vector $\op{x}(t)\in\mathbb{R}^N$ whose components are $x(t)_{i} = I(t)_{i} / I_{tot}(t)$.

The reproduction ratio $R$ of this process is the eigenvalue of $\op{R}$ with the largest absolute value (spectral radius)~\cite{diekmann_definition_1990,nishiura_pros_2010}, which is real, positive and nondegenerate (see Supplementary Methods Section 1.1). Its corresponding eigenvector $\op{v}$ is also strictly positive ($v_i>0$) we normalize it so that $\sum_i v_i=1$.

Measuring the reproduction ratio of the system thus requires knowledge of the spectral structure of $\op{R}$, i.e., of the spatial structure of the epidemic. We call this {\itshape reference} reproduction ratio as opposed to what surveillance instead estimates from the evolution of the incidence of infections or their proxies. 
This may happen globally, at the level of the entire population, or locally in each community.
In our framework, the population-level estimated reproduction ratio is $R^{estim}(t) = I_{tot}(t+1)/I_{tot}(t)$, i.e., the generational growth rate. The local community-level estimated reproduction ratio is instead $r^{estim}_i(t) = I(t+1)_i/I(t)_i$.

A simple observation then underpins our study: in general $R^{estim}(t)$ and $r^{estim}_i(t)$ may be different from $R$, the spectral radius of $\mathbf{R}$, and, if that is the case, surveillance will not estimate the reference reproduction ratio.

To explore this, we will first determine the conditions leading to an unbiased measure of the reproduction ratio: $R^{estim}(t)=R$.

\section*{When reference and estimated reproduction ratios match}
By virtue of the Perron-Frobenius theorem, $\op{R}^t\rightarrow R^t \op{v} \op{v}^*$ asymptotically at large $t$, where $\op{v}^*$ is the dual of $\op{v}$ (and left Perron eigenvector of $\op{R}$) and normalized so that $\op{v}^*\op{v}=1$.
Equation~\eqref{eq:Ievol} becomes $\op{I}(t)\rightarrow \left[\op{v}^* \op{I}(0)\right] R^t \op{v}$, which implies that $\op{x}(t)\rightarrow \op{v}$, $R^{estim}(t)\rightarrow R$ and $r^{estim}_i(t)\rightarrow R$.
The epidemic dynamics thus brings the spatial distribution of infections toward $\op{v}$, which we will refer to as the {\itshape equilibrium spatial distribution of infections}.
Thus, for any epidemic dynamics, if infections are spatially distributed as the equilibrium distribution ($\op{x} = \op{v}$), then the error is zero and the reference reproduction ratio is measured both globally $(R^{estim}=R)$ and locally $r^{estim}_i=R$.
The last equality implies the existence of a detailed balance between locally generated, imported and exported infections in every community (see Supplementary Methods Section 1.2).
Fig.~\ref{fig:fig1} shows evidence of the convergence to $\op{v}$ during the COVID-19 epidemic in France in late 2020 and early 2021.
We reconstructed $\op{R}$ from mobility data from Meta~\cite{Colocation_Maps}, a multinational technology company, and $\op{x}$ from hospitalization data (see Fig.~\ref{fig:fig1}, \nameref{sec:Rfromdata} and Supplementary Methods Section 1.3).
In a period when $\op{R}$ was fairly constant the angle between $\op{x}$ and $\op{v}$ consistently decreased, a sign that $\op{x}$ was evolving towards the equilibrium distribution, as the theory predicts.

Finally, there is a class of operators $\op{R}$ for which the reference and population-level estimated reproduction ratios match at all times and even out of equilibrium.
Indeed, if $\rho_i\equiv\sum_j R_{ji}$, the expected total number of secondary infections, is the same across communities, then $R^{estim}=\rho_i=R\,\forall i$ identically (see \nameref{sec:homo}). This implies that only the combination of spatial epidemic coupling and spatial heterogeneity in the transmission potential $\rho_i$ may cause $R^{estim}$ to be different from $R$. Notably, this is true only globally and $r^{estim}_i$ may be different from $R$ even if $\rho_i=R$.

\begin{figure}[h!]
	\centering
	\includegraphics[width=1\linewidth]{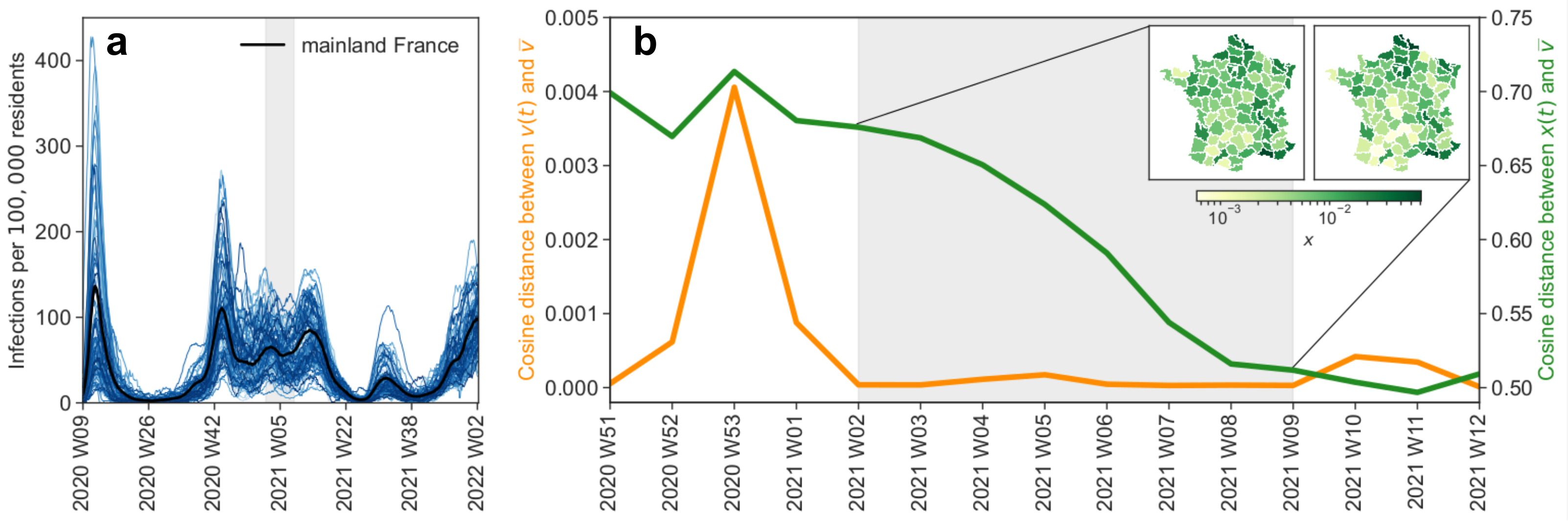}
	\caption{{\bfseries Convergence to the equilibrium spatial distribution of infections.}
 \textbf{a}, Estimate of the daily number of new COVID-19-related infections per 100,000 residents from week 9 of 2020 to week 3 of 2022 (Feb 23, 2020 to Jan 23, 2022). See Supplementary Methods Section 1.3 for how we estimated infections from hospitalizations and the comparison with serological surveys.
 Each curve is that of one of the 94 departments (administrative level 2) of mainland France excluding the region of Corsica, with a 14 days window rolling average. The period from week 2 to week 9 of 2021 (Jan 11 to Mar 7), selected as a period in which $\op{v}$ was fairly constant, is highlighted in grey both in \textbf{a} and \textbf{b}.
 The small dept. Territoire de Belfort ($0.2\%$ of the French population) is excluded due to a miscounting of hospitalizations coming from neighboring departments by the health authorities - see Supplementary Fig.~S3 and S4.
 \textbf{b}, The orange curve (left $y$ axis) reports the weekly cosine distance between $\op{v}$ (computed from mobility data from that week - see \nameref{sec:Rfromdata}) and its average ($\bar{\op{v}}$) on the time interval from week 51 of 2020 to week 12 of 2021 (see Supplementary Fig.~S13). The cosine distance between $\op{v}$ and $\bar{\op{v}}$ is defined as $1-\op{v}\cdot\bar{\op{v}} / \left( \|\op{v}\| \|\bar{\op{v}}\| \right)$. The green curve (right $y$ axis) reports the cosine distance between $\bar{\op{v}}$ and the weekly spatial distribution of COVID-19 estimated infections $\op{x}$ in French departments. The two inset maps show the spatial distribution $\op{x}$ in week 2 and week 9 of 2021, respectively (edges of the time interval in gray).
 }\label{fig:fig1}
\end{figure}

\section*{When reference and estimated reproduction ratios do not match}

\begin{figure}[h!]
	\centering
	\includegraphics[width=.9\textwidth]{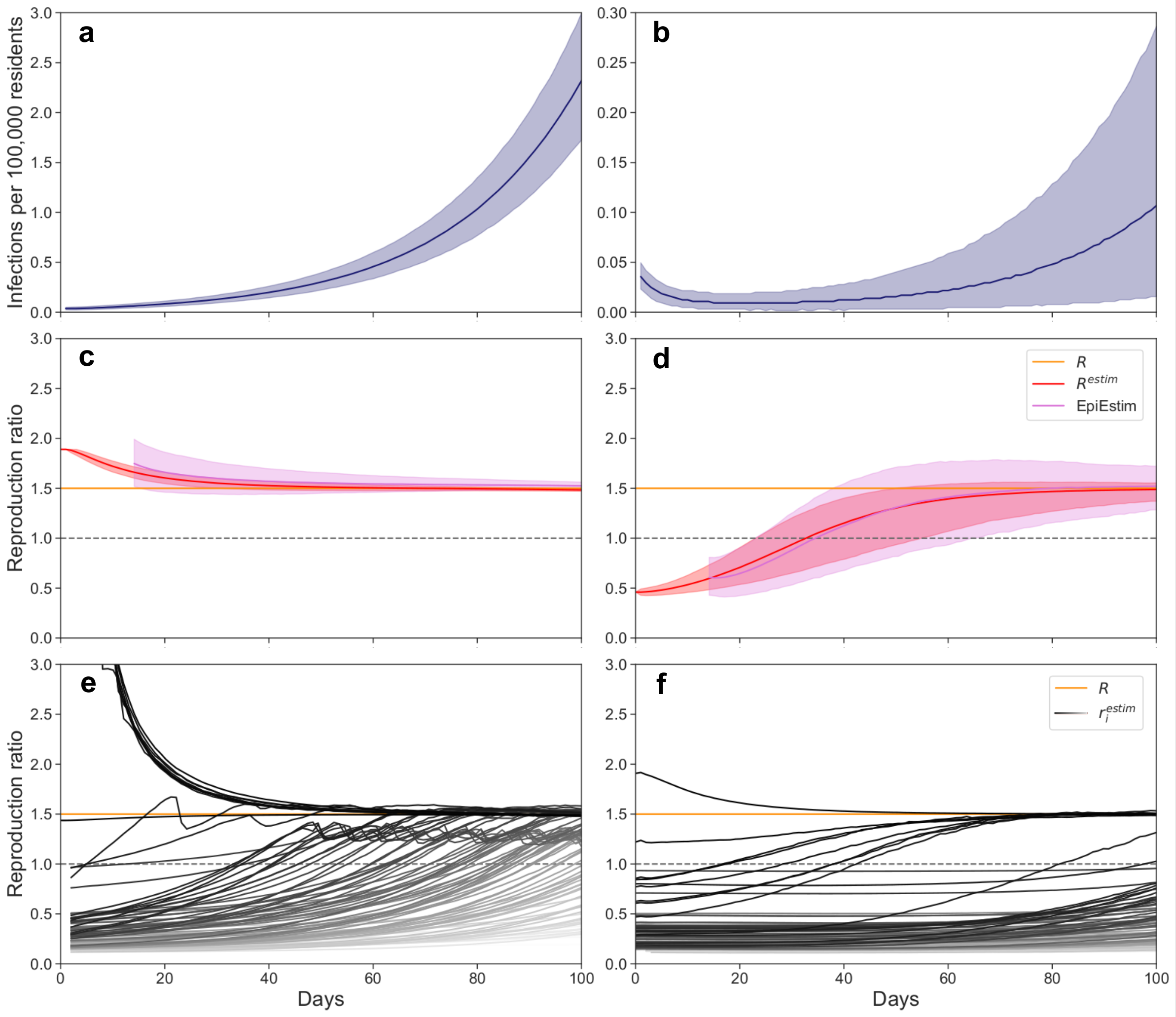}
	\caption{\label{fig:fig2}{\bfseries Comparison of the reference and measured reproduction ratios.} We generated stochastic simulations of a COVID-19-like epidemic in France using a metapopulation model, whereby we divided the French population in 94 spatial communities corresponding to the department of mainland France minus Corsica, and modeled within- and between-community mixing using mobility data by Meta. See \nameref{sec:metapop} and Supplementary Methods Section 1.4 for details on how we built $\mathbf{R}$ from mobility data and on the epidemic spread model. Two initial conditions are tested: in \textbf{a}, \textbf{c} and \textbf{e} $100$ initial infections are seeded in the department of Paris, while in \textbf{b}, \textbf{d} and \textbf{f} $100$ initial infections are seeded among departments proportionally to their population. \textbf{a} and \textbf{b} show the daily number of incident infections per 100,000 individuals. $R$ (yellow) is the reference reproduction ratio, $R^{estim}$ (red) is the estimated reproduction ratio as described in the paper, $r_i^{estim}$ are the locally estimated reproduction ratios, and {\itshape EpiEstim} (purple) is the reproduction ratio computed from incidence data using the {\itshape EpiEstim} package~\cite{cori_new_2013} (see Supplementary Methods Section 1.4 for details). We run $10,000$ runs of the stochastic model, and plot medians with solid lines and $95\%$ confidence intervals with shaded areas. For the $r_i^{estim}$, confidence intervals are omitted for readability.}
\end{figure}

We now focus on the out-of-equilibrium dynamics ($\op{x}(t)\not=\op{v}$) and measure the bias on $R$ as the relative difference between the estimated and the reference reproduction ratios
\begin{equation}
    \Delta(t) = \frac{R^{estim}(t)-R}{R}. 
    \label{eq:delta1}
\end{equation}
We call $\Lambda_\alpha$ ($\alpha=1,\cdots,N-1$) the (possibly degenerate) eigenvalues of $\op{R}$ other than $R$ and, by Perron-Frobenius theorem, $|\Lambda_\alpha|<R$.
With calculations reported in \nameref{sec:delta_calc}, we find that 
\begin{equation}
    \Delta(t) = C(t) \sum_\alpha z_\alpha \left(1-\frac{\Lambda_\alpha}{R}\right) \left( \frac{\Lambda_\alpha}{R} \right)^t,
    \label{eq:delta2}
\end{equation}
where $C(t)$ is positive and asymptotically constant, and $z_\alpha$ is a (possibly complex) number proportional to the projection of the initial condition $\op{x}(0)$ on the $\alpha$-th mode.
The modes in equation~\eqref{eq:delta2} for which $\Lambda_\alpha\approx R$, or that are almost orthogonal to the initial configuration $\op{x}(0)$, are suppressed from the start and do not bias the estimate of the reproduction ratio.
The other modes, instead, possibly do, with an effect that becomes smaller with a characteristic decay time $\tau_{\alpha}=1/\log\left(R/|\Lambda_\alpha|\right)$.
In addition, those modes for which $\Lambda_\alpha$ is not real and positive oscillate with period $T_{\alpha}=2\pi/|\theta_\alpha|$, where $\theta_\alpha=\arg \Lambda_\alpha$ (with $\theta_\alpha\in(-\pi,\pi]$) --~see \nameref{sec:delta_calc_tau_T}.
These oscillations will be visible if faster than their characteristic decay time: $T_\alpha\leq \tau_\alpha$. This gives the inequality
\begin{equation}
  \frac{|\Lambda_\alpha|}{R} \geq e^{-\frac{|\theta_\alpha|}{2\pi} } \geq e^{-\frac{1}{2}} \approx 0.61.
  \label{eq:oscillation_visible}
\end{equation}
The strict equality occurs when $\Lambda_\alpha$ is real and negative ($\theta_\alpha=\pi$).
Furthermore, using the same approach, we prove in Supplementary Methods Section 1.2 that oscillations may be visible in the local estimated reproduction ratios even when the global $\Delta(t)$ never changes sign (see Fig.~\ref{fig:fig2}\textbf{e}).

To test the predictions of our theory in a realistic scenario, we considered again the COVID-19 epidemic in France and built a stochastic metapopulation model using the same data as in Fig.~\ref{fig:fig1}\textbf{b}.
The details of the model are reported in \nameref{sec:metapop} and Supplementary Methods Section 1.4. We measured the reference and the estimated reproduction ratios, reported in Fig.~\ref{fig:fig2}, which shows that surveillance-based estimates may remain consistently biased for a long period and, depending on where the epidemic wave started , they may either overestimate or underestimate the reference reproduction ratio (see Supplementary Fig.~S9 for more initial conditions tested). The case depicted in Fig.~\ref{fig:fig2}\textbf{d} is particularly concerning: during the first month of the simulated epidemic, surveillance records a lower-than-one reproduction ratio which indicates decreasing incidence (visible in Fig.~\ref{fig:fig2}\textbf{b}), mistakenly pointing to a subsiding outbreak. In reality, the reproduction ratio is well above one, and only after two months of simulated epidemic does the surveillance based estimate reach this value.
Alongside the estimate of $R^{estim}$ given within the framework of the Galton-Watson process, in Fig.~\ref{fig:fig2}\textbf{c},\textbf{d} we also provide an estimate of the reproduction ratio by feeding incident cases to the library {\itshape EpiEstim}~\cite{cori_new_2013}, a popular tool to compute the reproduction ratio from surveillance data. The fact that the two measures overlap confirms that the Galton-Watson process correctly reproduces the phenomenology under study even in realistic scenarios.
Notwithstanding, more detailed frameworks ~\cite{diekmann_definition_1990,white_estimating_2013,trevisin_spatially_2023} could be used to study the impact of heterogeneous generation intervals.

Finally, Fig.~\ref{fig:fig2}\textbf{e} and Fig.~\ref{fig:fig2}\textbf{f} show that locally measured reproduction ratios converge to the reference value at different times and with different speeds and that, at the same moment in time, some communities overestimate $R$ and some underestimate it (analytical proof in \nameref{sec:overunder}.

Fig.~\ref{fig:fig2} shows no oscillations in the sign of $\Delta(t)$, compatible with the fact that the reconstructed operator $\op{R}$ has only real and positive eigenvalues. We extended our analysis to 32 European countries and none showed oscillations in $\Delta(t)$ at the considered spatial resolution (admin-2).
We discuss the details of this analysis and the possible realistic conditions leading to oscillations in Supplementary Methods Section 1.5 and Supplementary Fig.~S1.

In the absence of oscillations, the estimated reproduction ratio consistently either overestimates or underestimates the reference reproduction ratio, as $\Delta(t)$ decays to zero without ever changing sign. In this case, we can determine the sign of the bias from the initial condition: $\Delta(0)=\sum_j \rho_j x(0)_j-R$. By the Perron-Frobenius theorem, $j_{min},j_{max}$ exist so that $\rho_{j_{min}}\leq R$ and $\rho_{j_{max}}\geq R$. Thus, the initial location of infections will completely determine the sign of the error that surveillance will make. If the epidemic starts $j_{max}$ --~or in general in communities with high transmission potential~-- surveillance will consistently overestimate the reference reproduction ratio until the bias decays to zero. Conversely, if it starts in $j_{min}$ --~or in communities with low transmission potential~-- surveillance will underestimate $R$.
The spatial structure of the epidemic thus generates something similar to a {\itshape core group effect}, an epidemiological concept whereby the disease is initially confined to specific groups where strong clustering, high susceptibility or specific behavioral traits result in transmission rates higher than in the general population --~like, for instance, schoolchildren and influenza~\cite{nishiura_transmission_2009}~-- and requires specific corrections to epidemic models~\cite{bansal_when_2007,nishiura_pros_2010,chowell_characterizing_2016,liu_measurability_2018}.
In our case, areas with higher-than-average $\rho_i$ play the role of core groups, skewing the estimated reproduction ratio to values higher than $R$. We restate, however, that if the epidemic is instead localized in areas with lower-than-average $\rho_i$ the effect is opposite and leads to an underestimate of $R$, something that is not typically considered in the standard core-group theory.
And just as core group effects may make it harder to assess and predict the population-level impact of an epidemic, the bias that the spatial structure induces on the reproduction ratio prevents an accurate and timely estimate of the epidemic trajectory, possibly misinforming public health response.
The next section will present a method to overcome this.
We remark, however, that, at present, our formalism covers only core-groups effects related to the spatial structure. Other drivers, like the presence of clusters in the contact network, require additional complexity: from models featuring an initial subexponential growth of infections~\cite{chowell_characterizing_2016}, to some including the explicit contact network~\cite{liu_measurability_2018} inside each spatial community. Both can be future extensions of the model presented here, but they require changing the functional form of equation~\eqref{eq:Ievol} in the former case, or forgoing the analytical approach altogether in the latter case.
Another known source of bias on the reproduction ratio is the importation of infections from outside of the system, e.g, from other countries~\cite{nishiura_estimation_2010}.
Corrections exist for the initial overestimation coming from miscounting imported infections as infections generated within the systems, and could be used in combination with ours~\cite{mercer_effective_2011}.
Our correction, however, also addresses the impact that imported infections may have in changing the spatial distribution of the epidemic. Indeed if importations are distributed differently from the equilibrium distribution $\op{v}$, and there is no reason they should, they will enhance the bias on the reproduction ratio arising from $\op{x}$ being different from $\op{v}$, which is the object of our study.

\section*{Correction to surveillance data}

\begin{figure}[h!]
	\centering
	\includegraphics[width=1\textwidth]{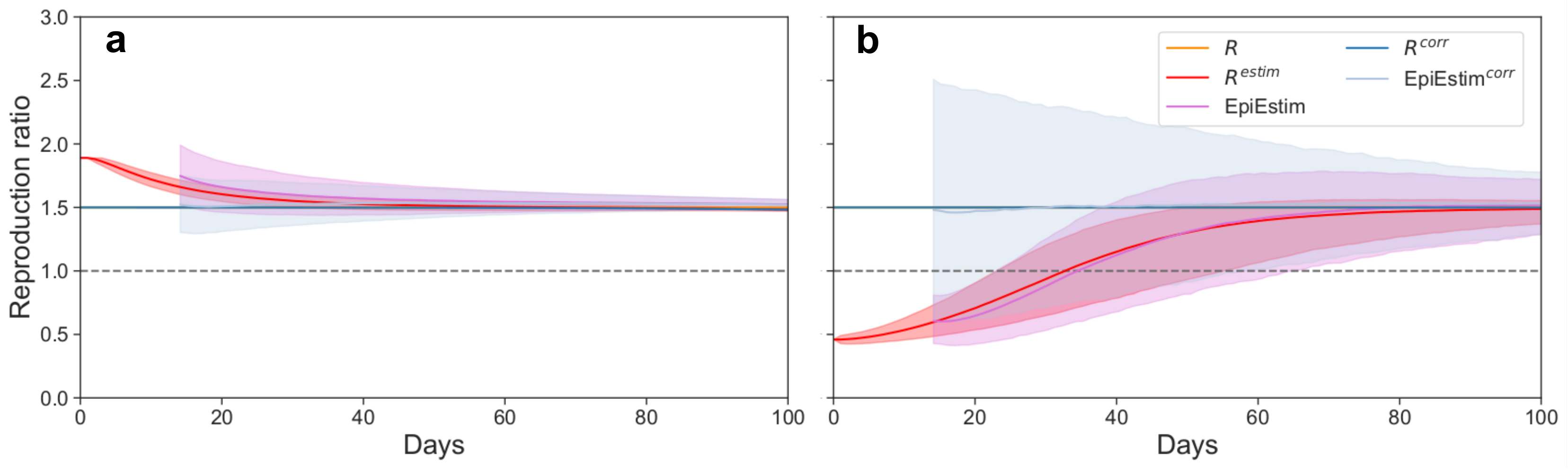}
	\caption{\label{fig:fig3}{\bfseries Corrected reproduction ratio.}
 This figure uses the same stochastic epidemic model as in Fig.~\ref{fig:fig2} and the same initial conditions, \textit{i.e.}, panels \textbf{a} and \textbf{b} correspond to panels \textbf{c} and \textbf{d} in Fig.~\ref{fig:fig2} respectively. Medians and $95\%$ confidence intervals are computed and shown over $10,000$ runs. We compare the standard reproduction ratio measured from surveillance data --~both in the Galton-Watson formalism (red) and with \textit{EpiEstim} (purple) on incidence data~-- and the corresponding $\op{v}^*$-corrected estimates: Galton-Watson formalism (blue) and with $\textit{EpiEstim}$ (light blue) on modified incidence data $I^{(v)}_{tot}$.
 }
\end{figure}

So far we have proven that surveillance-based estimates of the reproduction ratio may be biased.
We will now propose a way to correct for this bias.
The global incidence of infections, used to estimate the reproduction ratio, is trivially proportional to the unweighted average of the incidence across communities: $I_{tot}(t) = N \left( \sum_i I_i(t) / N \right)$.
We now define a modified incidence that weighs local infections with the Perron dual vector:
\begin{equation}
 I^{(v)}_{tot}(t) = N \left( \frac{\sum_i v^*_i I_i(t)}{\sum_i v^*_i} \right) = N\sum_i v^*_i I_i(t) = N \op{v}^* \op{I}(t).
 \label{eq:Iv}
\end{equation}
Then, we use it to get a new estimate of the reproduction ratio:
\begin{equation}
    R^{corr}(t) = \frac{ I^{(v)}_{tot}(t+1) }{ I^{(v)}_{tot}(t) } = \frac{\op{v}^* \op{I}(t+1)}{\op{v}^* \op{I}(t)} = \frac{\op{v}^* \op{R}\op{I}(t)}{\op{v}^* \op{I}(t)} = R \frac{\op{v}^* \op{I}(t)}{\op{v}^* \op{I}(t)} = R,
\end{equation}
and find it is identically equal to the reference value.
The practical advantage for epidemic monitoring is clear: our correction gives an unbiased estimate of the reproduction ratio from surveillance data all along the epidemic wave, unlike traditional measures.
It has, however, two potential drawbacks.
The former is that if the outbreak starts in communities where $v^*_i$ is small, then $\op{v}^* \op{I}(t)$ will be very small: stochastic fluctuations would then cause large changes in $R^{corr}$. In that case then $R^{corr}$ may well be accurate, but not precise.
However, no initial condition can be orthogonal to $\op{v}^*$ whose entries are strictly positive, so even if $\op{v}^*\op{x}$ is initially small, it is likely to increase quickly and with it the precision of the measurement.
In Fig.~\ref{fig:fig3} we show that $R^{corr}$ accurately measures the reference reproduction ratio from the beginning of the epidemic wave, in the case of the simulated epidemics of Fig.~\ref{fig:fig2}. Notably, Fig.~\ref{fig:fig3} also shows that if you feed $I^{(v)}_{tot}(t)$ to {\itshape EpiEstim} instead of $I_{tot}(t)$ you will also completely remove the bias on the estimate of the reproduction ratio. Our proposed modified incidence can then be readily incorporated to standard tools for public health surveillance, to improve their accuracy.

\begin{figure}[h!]
	\centering
	\includegraphics[width=0.7\textwidth]{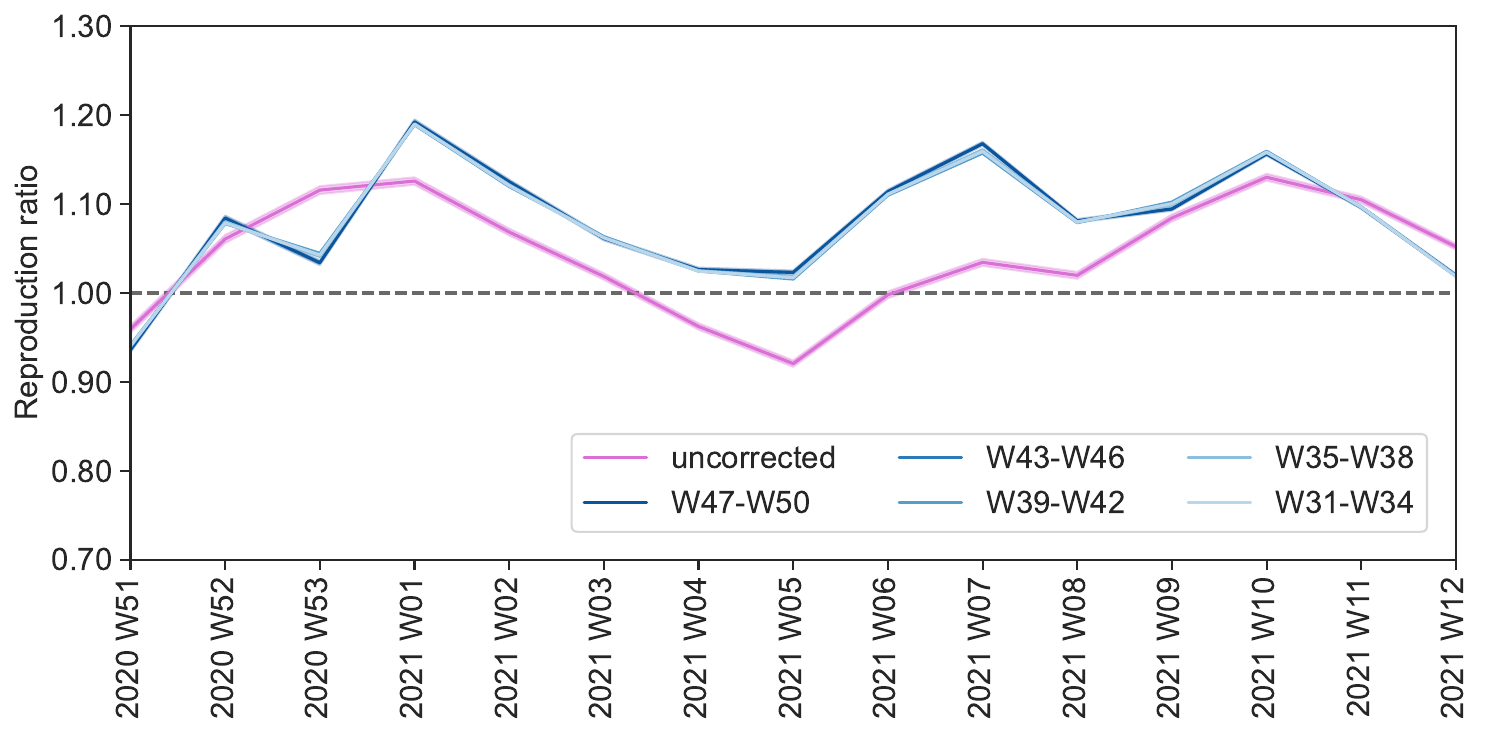}
	\caption{\label{fig:fig4}{\bfseries Application of the proposed correction to COVID-19 data in France.} The pink curve reports the reproduction ratio of COVID-19 in France obtained from country-level incidence, in the same time interval as in Fig.~\ref{fig:fig1}\textbf{b}, using the package \textit{EpiEstim}. The blue curves reports the corrected reproduction ratio obtained by feeding \textit{EpiEstim} the corrected incidence as in equation~\eqref{eq:Iv}. For this, the Perron dual vector $\op{v}^*$ is reconstructed from mobility data at different times prior the observation period, specified in the legend as week ranges of year 2020. The shaded areas are the 95\% credibility intervals computed by \textit{EpiEstim}. For readability, only those of the pink curve and the darkest blue curve are shown. The different blue curves are hardly distinguishable because they overlap. See \nameref{sec:Rfromdata} for a detailed explanation on how to compute $\op{v}^*$ from data.}
\end{figure}

The latter potential drawback is that our correction requires knowing $\op{v}^*$.
We argue, however, that this does not require knowing or measuring $\op{R}$ in real time and that
a good estimate of $\op{v}^*$ for epidemic monitoring can be computed during {\itshape peace time}, from past population and mobility data (pre-epidemic, or from data collected during earlier epidemic phases). Indeed $\op{v}^*$ is more stable than $\op{R}$ for the fact that any change happening homogeneously across communities (e.g., changes in the rate of immunity, public health interventions) would change the latter, not the former.
Fig.~\ref{fig:fig4} compares the standard estimated reproduction ratio of COVID-19 in France between late 2020 and March 2021 to our correction.
The former is computed with EpiEstim on inferred case incidence, the latter is computed with EpiEstim on the corrected incidence $I^{(v)}_{tot}$, with $\op{v}^*$ computed from various samples of past mobility data, all yielding stable estimates.
Our correction seems to point to the fact that traditional surveillance underestimated the reproduction ratio of COVID-19 in France during January and February 2021.
Notably, surveillance recorded a lower-than-one reproduction ratio during more than two weeks (see also official reports from that time~\cite{sante_publique_france_covid-19_nodate}), indicating a subsiding epidemic wave. This is at odds with what we know happened: a growing epidemic wave --~the French {\itshape third wave}~-- that led to a national lockdown, enforced on April 3 2021, i.e., immediately after the time window depicted in Fig.~\ref{fig:fig4}. Our corrected reproduction ratio would have instead consistently signaled a growing epidemic wave throughout the first three months of 2021m by correctly weighing the infections in departments with low transmission potential --~see Supplementary Fig. S12 and S13.
This discrepancy carries great significance when put into the context of the debate over public health response at that time. In early 2021 a national curfew was in effect but cases were rising due to the introduction and gradual takeover of the Alpha variant of SARS-CoV-2.
Authorities were wary of additional restrictions and were relying on mass vaccination despite models suggesting that it might not be enough~\cite{di_domenico_impact_2021} --~only $3\%$ of the population had received at least one dose by mid February~\cite{sante_publique_france_covid-19_nodate} (week 6 of 2021 in Fig.~\ref{fig:fig4}).
It is conceivable, albeit circumstantial, that the fact that surveillance underestimated the severity of the wave could have contributed to delaying the enforcement of stricter movement restrictions, which became anyway inevitable later in April.

We remark that our method removes a source of bias that it is inherent to the spatial dynamics of the system and may be present even when surveillance data are accurate, representative and unbiased. In this sense it should integrate methodologies that address other sources of bias such as uneven susceptibility and case detection rate, reporting delays, inaccurate reconstruction of infections from proxy timeseries, inaccurate estimation of the generation time~\cite{nishiura_correcting_2010,liu_measurability_2018,gostic_practical_2020,sherratt_exploring_2021,jorge_estimating_2022}: We prove the computational feasibility of this integration in our library~\cite{library}.
It is also important to stress, however, that our correction is as good as the data used to calculate it, and specifically the data used to reconstruct $\op{R}$.
The different global health threats we identified in the introduction entail different minimum requirements on the data: COVID-19 and influenza need data on local and between-community mixing, arboviral diseases (dengue, chikungunya, Zika) data on where residents of each community spend their time and the relative spatial distribution of the vector population (Aedes mosquitoes).
Finally, spreading routes and disease natural history will inform the specific data needs in the case of newly emerging pathogens or strains.
Overall, however, large-scale estimates of human mobility and mixing will be necessary, and they increasingly come from mobile phone data.
The impact of their own limitations~\cite{wesolowski_impact_2013,tizzoni_use_2014,wesolowski_connecting_2016,lai_exploring_2019,sekara_mobile_2019} will require case-by-case assessment when these data are used to compute our correction.
In our case study, Fig.~\ref{fig:fig4} shows that the correction is robust to choosing human mixing data from different time periods in our case study.
Also, both the data provider and we implemented adjustments to improve the accuracy and representativeness of the colocation data (see Ref.~\cite{Colocation_Maps} and \nameref{sec:Rfromdata}).
Finally, our correction is currently static and works over periods of fairly constant $\op{R}$: future studies should address how $\op{v}^*$ should change in time to consistently correct for the bias on the reproduction ratio, by explicitly coupling population mixing and disease natural history and handling exogenous shifts in $\op{R}$ due to seasonal trends, behavioral changes and interventions.
    
Our study describes a practicable way to improve the accuracy of the information that flows from epidemiological surveillance to public health policymakers. And better information may lead to more effective policies for preventing and controlling epidemic threats.


\newpage



\section*{Methods}

\subsection*{Homogeneous transmission potential}
\label{sec:homo}

Let us rewrite the estimated reproduction ratio in matrix form as
\begin{equation}
    R^{estim}(t) = \frac{I_{tot}(t+1)}{I_{tot}(t)} = \op{F}^T \op{R} \op{x}(t),
    \label{eq:S}
\end{equation}
where we introduced $\op{F}$ as the unit column vector ($F_i=1\,\forall i$). If we assume that $\op{v}^*=\op{F}^T$ ($v^*_i=1$) then we can apply $\op{R}$ leftwards in equation~\eqref{eq:S} and get $R^{estim}(t)=R$ at any time and for any spatial distribution $\op{x}$. Now, the requirement $\op{v}^*=\op{F}^T$ imposes that $\op{R}$ is proportional to a left-stochastic matrix: indeed $\op{F}^T \op{R}=R \op{F}^T$ means $\rho_i=\sum_j R_{ji}=R$, so that each column sums to $R$.

\subsection*{Calculation of $\Delta(t)$: proof of equation~\eqref{eq:delta2}}\label{sec:delta_calc}

Combining equation~\eqref{eq:Ievol} and equation~\eqref{eq:S} we get the time evolution of the estimated reproduction ratio:
\begin{equation}
    R^{estim}(t) = \frac{ \op{F}^T \op{R}^{t+1}\op{x}(0) }{ \op{F}^T \op{R}^{t}\op{x}(0)}.
\end{equation}
We insert this into equation~\eqref{eq:delta1} and get
\begin{equation}
    \Delta(t) = \frac{1}{R} \frac{ \op{F}^T (\op{R}-R)\op{R}^{t}\op{x}(0) }{ \op{F}^T \op{R}^{t}\op{x}(0) }.
    \label{eq:delta3}
\end{equation}
We introduce the eigenvectors of $\op{R}$ (other than $\op{v}$): $\op{w}_\alpha$ eigenvector with corresponding eigenvalue $\Lambda_\alpha$. Analogously we define the corresponding dual vector $\op{w}^*_\alpha$. Then, we decompose $\op{F}^T$ in the dual basis: $\op{F}^T = \op{v}^* + \sum_\alpha \left( \op{F}^T \op{w}_\alpha \right) \op{w}^*_\alpha$. Using this decomposition in equation~\eqref{eq:delta3} and applying $\op{R}$ leftwards on the dual eigenvectors we get
\begin{equation}
    \Delta(t) = - \frac{ \op{F}^T\left[ \sum_\alpha \left(\frac{\Lambda_\alpha}{R}\right)^t \left( 
1-\frac{\Lambda_\alpha}{R} \right) \op{w}_\alpha\op{w}_\alpha^* \right]\op{x}(0) }{ \op{F}^T\left[ \op{v}\op{v}^* + \sum_\alpha \left(\frac{\Lambda_\alpha}{R}\right)^t \op{w}_\alpha\op{w}_\alpha^* \right]\op{x}(0) }.
\label{eq:delta4}
\end{equation}
The denominator is $C(t)$ in equation~\eqref{eq:delta2}:
\begin{equation}
    C(t) = \frac{ 1 }{ \op{F}^T\left[ \op{v}\op{v}^* + \sum_\alpha \left(\frac{\Lambda_\alpha}{R}\right)^t \op{w}_\alpha\op{w}_\alpha^* \right]\op{x}(0) }.
\end{equation}
$C(t)$ is always strictly positive because it is proportional to $\op{F}^T \op{R}^{t}\op{x}(0)$ and tends to $\op{v}^* \op{x}(0)$, i.e., the component of the initial condition onto the eigenspace of the Perron eigenvalue. This component is always nonzero because no $\op{x}(0)$ is nonnegative (as it is a spatial distribution of infections) and no nonnegative vector can be orthogonal to a strictly positive vector.
It is thus the numerator which gives the trend and sign of $\Delta(t)$. 
Equation~\eqref{eq:delta4} then gives the value of the factors $z_\alpha$ in equation~\eqref{eq:delta2}:
\begin{equation}
    z_\alpha = - \op{F}^T \left( \op{w}_\alpha \op{w}^*_\alpha \right)\op{x}(0)
\label{eq:z_alpha}
\end{equation}
In the case of degenerate eigenvalue one should simply replace $\op{w}_\alpha\op{w}^*_\alpha$ with the appropriate projector over the whole eigenspace. Note that, as discussed before, the denominator in equation~\eqref{eq:delta4} is always real and positive so any complex phase of $z_\alpha$ must arise from $\Lambda_\alpha$ and $\op{w}_\alpha\op{w}^*_\alpha$.

\subsection*{Calculation of $\Delta(t)$: $\tau_\alpha, T_\alpha$}\label{sec:delta_calc_tau_T}

We isolate in equation~\eqref{eq:delta2} the contribution of each mode $M_\alpha(t)$: $\Delta(t) = {C(t)}\sum_\alpha M_\alpha(t)$, where
\begin{equation}
    M_\alpha(t) = M_\alpha(0) \left( \frac{\Lambda_\alpha}{R} \right)^t = M_\alpha(0)\left( \frac{|\Lambda_\alpha| }{R} \right)^t e^{i\theta_\alpha t} = M_\alpha(0)e^{-t/\tau_\alpha} e^{i \theta_\alpha t},
\end{equation}
where we used the definition of $\tau_\alpha$ given in the main text. The decaying term with characteristic time $\tau_\alpha$ is visible.

If $\Lambda_\alpha$ is real and positive then $\theta_\alpha=0$ and the oscillating term vanishes. If $\Lambda_\alpha$ is real and negative then $\theta_\alpha=\pi$ and the oscillating term becomes an alternating sign: $e^{i \theta_\alpha t}=(-1)^t$. This is an oscillation with period $T_\alpha=2$, which is compatible with the definition of $T_\alpha$ given in the main text. Finally, if $\Lambda_\alpha\not\in\mathbb{R}$, then then $\bar{\Lambda}_\alpha$ is also an eigenvalue, where the bar denotes complex conjugation. We will call $\bar{\alpha}$ the index corresponding to that eigenvalue: $\Lambda_{\bar{\alpha}}=\bar{\Lambda}_\alpha$. Also, the projector over the eigenspace of $\Lambda_{\bar{\alpha}}$ is the elementwise complex conjugate of the projector over the eigenspace of $\Lambda_\alpha$, meaning that $z_{\bar{\alpha}}=\bar{z}_\alpha$, and thus $M_{\bar{\alpha}}=\bar{M}_\alpha(0)$. Then $\alpha, \bar{\alpha}$ contribute in pair, as follows:
\begin{align}
    M_\alpha(t)+M_{\bar{\alpha}}(t) &= e^{-t/\tau_\alpha} \left[ M_\alpha(0)e^{i\theta_\alpha t} + M_{\bar{\alpha}}(0)e^{-i\theta_\alpha t} \right] \nonumber \\
    &= 2e^{-t/\tau_\alpha} \left|M_\alpha(0)\right| \Re e^{i\theta_\alpha t + \phi_\alpha} =2e^{-t/\tau_\alpha} \cos\left( \frac{2\pi}{T_\alpha} t + \phi_\alpha \right).
\end{align}
Here we used the definition of $T_\alpha$ given in the main text, explicitly showing the emergence of the oscillating term with period $T_\alpha$.

\subsection*{Local over- and underestimation of $R$}
The Collatz-Wielandt inequalities tell us that, for any spatial distribution of infections $\op{x}$, $\min_{i|x_i\not=0} (\op{R}\op{x})_i/x_i\leq R$ and $\max_{i|x_i\not=0} (\op{R}\op{x})_i/x_i\geq R$. Given that $r^{estim}_i=(\op{R}\op{x})_i/x_i$, out of equilibrium there will always be at least one community overestimating the reference reproduction ratio ($r^{estim}_i>R$) and one underestimating it ($r^{estim}_i<R$).
\label{sec:overunder}

\subsection*{Reconstruction of the reproduction operator from data}\label{sec:Rfromdata}
The main data used for the reconstruction of reproduction operators for mainland France are Meta Colocation Maps, which are calculated from mobile phone data and show the proportion of time that residents of different communities spend in proximity\cite{Colocation_Maps}. More precisely, they give the probability $p_{ij}$ that a randomly chosen person that is resident of community $i$ and a randomly chosen person resident of community $j$ are both located in a same $600m \times 600m$ square, during a randomly chosen five-minutes time window, in a given week.
The data were provided at the resolution of departments (ADM 2). To reconstruct $\op{R}$ from these data, we assumed that the expected number of secondary infections generated among the residents of community $i$, by a case who
is resident of community $j$, is given by $R_{ij} = C p_{ij} n_i$, where $n_i$ is the population of spatial patch $i$, and $C$ is an overall transmissibility parameter. Notably, while the value of the spectral radius of $\op{R}$ clearly depends on $C$, the left and right Perron eigenvectors $\op{v}$ and $\op{v}^*$ do not, and depend solely on the data. To compute $\op{v},\op{v}^*$ from $\op{R}$ we computed its right ($\op{v}$) and left ($\op{v}^*$) eigenvectors associated to the largest eigenvalue, using the Python function \texttt{numpy.linalg.eig} of \texttt{numpy} version 1.12.
The diagonal elements $p_{ii}$ of the Colocation Maps quantify the mixing within each community.
In a previous study we found that they overestimate it in densely populated areas and when people spend a lot of time at home~\cite{mazzoli_projecting_2021}. Using that same methodology, we thus corrected those entries using Meta Movement Range Maps (see Data availability). Movement Range Maps give the daily fraction $sp_i$ of residents of community $i$ that do not leave their $600m \times 600m$ home tile. The probability $p^{(sp)}_{ii}$ to detect co-locations among people staying in their separate homes and thus not actually mixing is, in community $i$,
\begin{equation}
    p^{(sp)}_{ii}=\frac{(sp_i d_i A)(sp_i d_i A -1)m_i}{n_i (n_i-1)}\, , 
\end{equation}
where $d_i$ is the population density in community $i$, $A=0.36\;km^2$ is the area of a single tile, $m_i$ is the number of tiles occupied by community $i$ and $n_i$ is the resident population population. We then discount $p^{(sp)}_{ii}$ from $p_{ii}$ to remove spurious colocations. We show in Supplementary Fig.~S2 the magnitude of this for the central week of the gray period of Fig.~\ref{fig:fig1}. For most departments it ranges between $1\%$ and $10\%$, exceeding the latter value only in the densely populated area of Paris and its surroundings.

\subsection*{Epidemic simulations}\label{sec:metapop}
The model of epidemic spread used in simulations is a stochastic discrete-time metapopulation model whereby spatially distinct communities are linked through mobility\cite{Colizza_Vesp_,balcan_multiscale_2009_,gomez-gardenes_critical_2018,chang_mobility_2020}.
We use a synthetic population based on census data from the National Institute of Statistics and Economic Studies (INSEE) in France.
We divide this population in $94$ spatial communities corresponding to the departments of mainland France except Corsica. Meta Colocation Maps \cite{Colocation_Maps} and Movement Range Maps are used to reconstruct the coupling $p_{ij}$ between communities $i$ and $j$ and the within-community $i$ mixing $p_{ii}$.
We use a SEIR compartmental model with $\epsilon$ as the rate of the E-to-I transition, $\mu$ rate of the I-to-R transition and $\beta$ overall transmission rate. The values of $\epsilon,\mu$ come from Ref.~\cite{Parameters_} (COVID-19-like), and $\beta$ is set so that $R=1.5$.
Following Ref.~\cite{Colocation_Maps}, the rate of force-of-infection to which those in community $i$ are subjected is $\beta \sum_j p_{ij} I_j$, where $I_j$ is the number of infected in $j$.
 We use a discretization time step $\Delta d=1 day$.
 The Supplementary Methods Section 1.4. describes in detail each step of the stochastic simulation model. There, we also compute the reproduction ratio for our model according to the next generation matrix formalism \cite{Next_Gen_Matrix_}, obtaining
\begin{equation}
    NGM_{ij}=\frac{\beta}{\mu}p_{ij}n_i,
\end{equation}
and $R=\mbox{spectral rad}(NGM)$ -- see Supplementary Methods Section 1.4. Again, $n_i$ is the population of community $i$.


\section*{Data availability}
\label{sec:data}
Meta Colocation Maps, which were used to reconstruct reproduction operators and to infer between- and within-community mixing for stochastic simulations can be requested at \sloppy\url{https://dataforgood.facebook.com/dfg/tools/colocation-maps}. Meta Movement Range Maps used to correct within-community colocations are available at \sloppy\url{https://data.humdata.org/dataset/movement-range-maps}. Hospital admission data in France are available at \sloppy\url{https://www.data.gouv.fr}. French census data can be found at \sloppy\url{https://www.insee.fr}. French departments shapefiles are available at \sloppy\url{https://www.data.gouv.fr/en/datasets/carte-des-departements-2-1/}. All websites accessed November 2023.

\section*{Code availability}
\label{sec:codeAvail}
The code used in this study is available here: \sloppy\url{https://github.com/ev-modelers/birello-surveillance}. A continuously developed and maintained library allowing to implement our correction to estimates of the reproduction ratio from surveillance data can be found at \sloppy\url{https://github.com/ev-modelers/rt-from-surveillance}.

\section*{Acknowledgments}
Colocation data were available thanks to {\itshape Data For Good at Meta}.

\section*{Author contributions}
E.V. conceived and designed the study.
E.V. and M.R.F. developed the theory.
P.B. developed the code and the publicly available library, performed the numerical simulations and analyzed the results.
B.W. analyzed surveillance data and reconstructed infections from them.
P.B., M.R.F., B.W., V.C., E.V. interpreted the results.
E.V. drafted the Article.
P.B., M.R.F., B.W., V.C., E.V. contributed to and approved the final version of the Article.

\section*{Competing interests}
The authors declare no competing interests.

\newpage



\includepdf[pages=-]{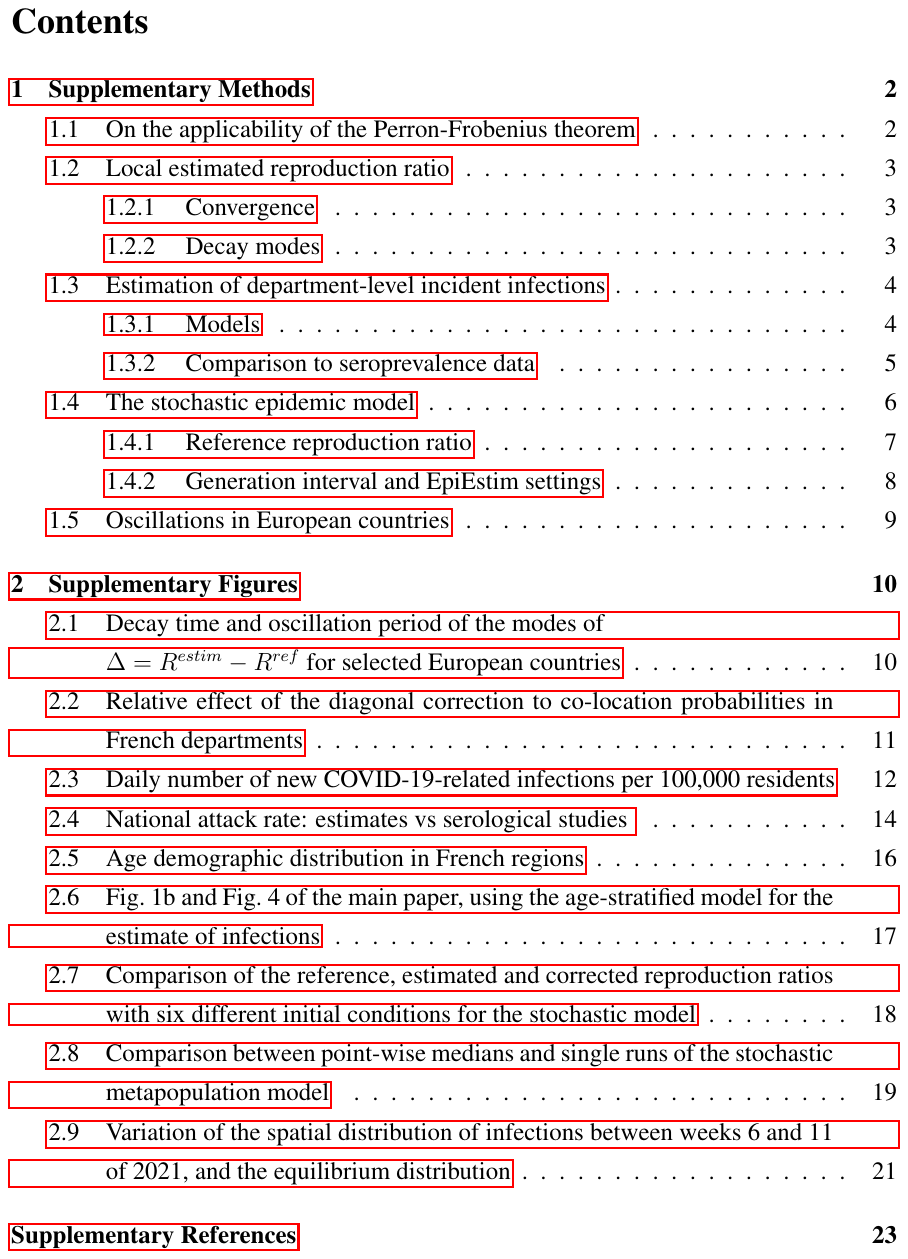}


\end{document}